%% file: sample-base.tex
\documentclass[manuscript,nonacm]{acmart}
\AtBeginDocument{%
  }

\setcopyright{acmlicensed}
\copyrightyear{2025}
\acmYear{2025}
\acmDOI{XXXXXXX.XXXXXXX}
\acmConference[Conference acronym 'XX]{Make sure to enter the correct
  conference title from your rights confirmation email}{June 03--05,
  2025}{Woodstock, NY}

\acmISBN{978-1-4503-XXXX-X/2018/06}

\usepackage{graphicx}
\usepackage[most]{tcolorbox}
\tcbset{enhanced, breakable, colback=white, colframe=black, boxrule=0.4pt}

\usepackage{xcolor}
\usepackage{capt-of}
\usepackage{setspace}
\usepackage{pifont}

\begin{document}

\title{Designing and Evaluating a Conversational Agent for Early Diagnosis of Alzheimer’s Disease and Related Dementias}

\author{Andrew G. Breithaupt}
\authornote{Both authors contributed equally to this research.}
\authornote{Department of Neurology}
\email{abreith@emory.edu}
\author{Nayoung Choi}
\authornotemark[1]
\authornote{Department of Computer Science}
\email{nayoung.choi@emory.edu}
\affiliation{%
  \institution{Emory University}
  \department{Department of Neurology}
  \department{Department of Computer Science}
  \city{Atlanta}
  \state{Georgia}
  \country{USA}
}

\author{James D. Finch}
\authornotemark[3]
\email{jdfinch@emory.edu}
\affiliation{%
  \institution{Emory University}
  \department{Department of Computer Science}
  \city{Atlanta}
  \state{Georgia}
  \country{USA}
}

\author{Jeanne M. Powell}
\authornotemark[2]
\email{jmpowe6@emory.edu}
\affiliation{%
  \institution{Emory University}
  \department{Department of Neurology}
  \city{Atlanta}
  \state{Georgia}
  \country{USA}
}

\author{Arin L. Nelson}
\authornotemark[2]
\email{arin.nelson@emory.edu}
\affiliation{%
  \institution{Emory University}
  \department{Department of Neurology}
  \city{Atlanta}
  \state{Georgia}
  \country{USA}
}

\author{Oz A. Alon}
\email{oalon@emory.edu}
\affiliation{%
  \institution{Emory University}
  \department{Emory College of Arts and Sciences}
  \city{Atlanta}
  \state{Georgia}
  \country{USA}
}

\author{Howard J. Rosen}
\authornotemark[2]
\email{howie.rosen@ucsf.edu}
\affiliation{%
  \institution{University of California, San Francisco}
  \department{Department of Neurology}
  \city{San Francisco}
  \state{California}
  \country{USA}
}

\author{Jinho D. Choi}
\authornotemark[3]
\email{jinho.choi@emory.edu}
\affiliation{%
  \institution{Emory University}
  \department{Department of Computer Science}
  \city{Atlanta}
  \state{Georgia}
  \country{USA}
}

\renewcommand{\shortauthors}{Breithaupt \& Choi et al.}

\begin{abstract}
\input{section/abstract}
\end{abstract}



\begin{teaserfigure}
  \includegraphics[width=\textwidth]{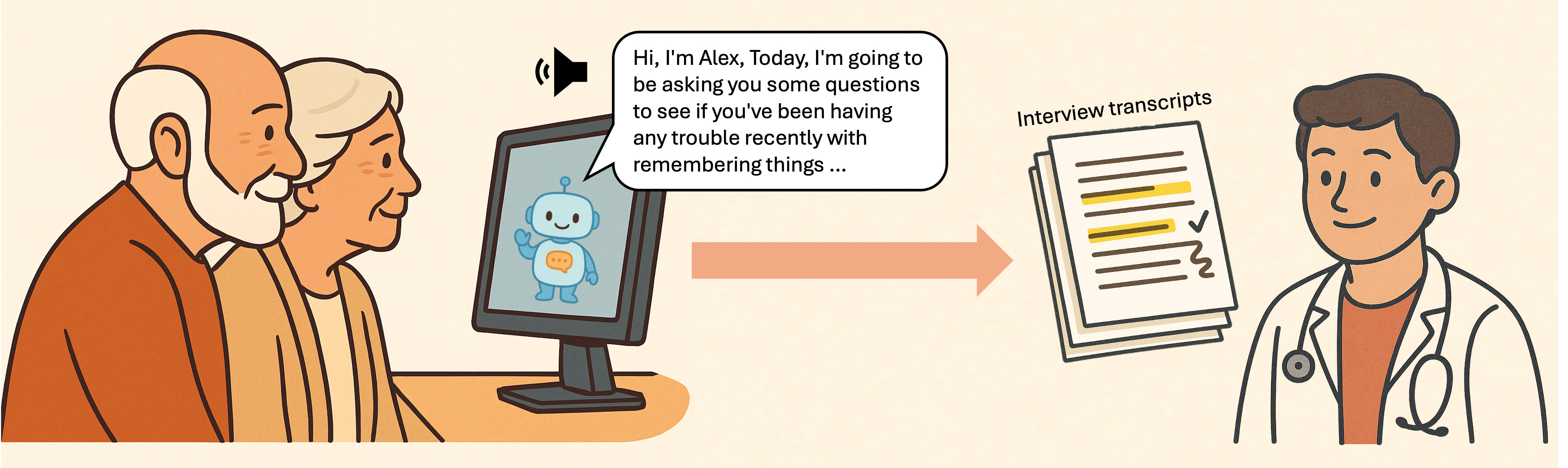}
  \caption{Overview of our voice-interactive conversational agent for ADRD diagnosis support. A patient and an informant (e.g., spouse) interact with the agent, and clinicians subsequently review the transcript to support diagnostic decision-making.}
  \label{fig:teaser}
\end{teaserfigure}

\received{19 January 2026}

\maketitle

\section{Introduction}
\input{section/introduction}

\section{Related Work}
\input{section/related_work}

\section{Method}
\input{section/method}

\section{Result}
\input{section/result}

\section{Discussion}
\input{section/discussion}

\section{Conclusion}
\input{section/conclusion}

\begin{acks}
We sincerely thank the patients and their families for their participation in this study.
\end{acks}

\bibliographystyle{ACM-Reference-Format}
\bibliography{sample-base}

\appendix
\input{section/appendix}

\end{document}

%% file: section/abstract.tex
Early diagnosis of Alzheimer's disease and related dementias (ADRD) is critical for timely intervention, yet most diagnoses are delayed until advanced stages. While comprehensive patient narratives are essential for accurate diagnosis, prior work has largely focused on screening studies that classify cognitive status from interactions rather than supporting the diagnostic process. We designed voice-interactive conversational agents, leveraging large language models (LLMs), to elicit narratives relevant to ADRD from patients and informants. We evaluated the agent with 30 adults with suspected ADRD through conversation analysis, user surveys, and analysis of symptom elicitation compared to blinded specialist interviews. Symptoms detected by the agent showed promising agreement with those identified by specialists. Users appreciated the agent’s patience and systematic questioning, which supported engagement and expression of complex, hard-to-describe experiences. While these findings suggest potential for conversational agents as structured diagnostic support tools, further validation with larger samples and assessment of clinical utility is needed before deployment.

%% file: section/introduction.tex
\label{sec:intro}

Alzheimer's disease and related dementias (ADRD) are rapidly becoming one of the most pressing global health challenges as populations age \cite{who2025facts, alz2024facts}. In the United States, the lifetime risk of dementia reaches 48\% for women and 35\% for men after age 55, with projected annual new cases doubling to more than 1 million by 2060 \cite{fang2025natmed}. Early diagnosis is crucial for accessing quality care, eligibility for new and emerging disease modifying therapies (DMT) and particularly for enabling research aimed to prevent dementia; however, over 50\% of dementia diagnoses are delayed until moderate or advanced stages in primary care \cite{Bradford2009Missed, Kotagal2015Factors} with greater delays among racial and ethnic minorities \cite{Babulal2019Disparities}. 
A severe shortage of specialists, combined with time and knowledge constraints especially in primary care settings, creates a growing gap between patient needs and available care \cite{Boustani2005JGIM, Bernstein2019JGIM, Liu2024RAND}.

The path to an accurate diagnosis starts with a comprehensive patient narrative (a "patient history"), which provides essential context for interpreting biomarkers, cognitive tests, and other diagnostic data \cite{Bouwman2022CSF, Dubois2024ClinicalBiologicalConstruct}. While memory care clinics at academic medical centers often provide 1 to 2 hours for rich patient–clinician interaction, primary care visits of 20 minutes or less rarely allow such extended dialogue or exploration of concerns \cite{Linzer2015}. Furthermore, with the average wait time for memory care clinics exceeding a year with projections to exceed 40 months by 2027, \cite{Liu2024RAND}, specialty clinics are also pressed to reduce visits time to adequately address the increasing demand. This challenge is particularly acute for older adults in under-resourced or rural settings, where specialist availability is limited \cite{liu2024geographic, giebel2025geographies}, making inclusive access to supportive, patient-centered interaction even more critical. These pressures also highlight the need for technology-mediated approaches that can extend such interactions beyond the constraints of clinic visits.

To address this gap, previous research has explored AI technologies in health screening contexts, including cognitive assessments for older adults \cite{TakeshigeAmano2024Digital, Yoshii2023ScreeningMCIHumanoidRobots, pahar2025cognospeak, deArribaPerez2022entertainmentbot}. These screening studies have primarily examined whether cognitive status (i.e., dementia, mild cognitive impairment (MCI), and healthy controls) can be accurately classified from AI-elicited interactions, rather than supporting clinicians in finding the cause of a patient's MCI or dementia. While better detection is certainly needed, such approaches in isolation add to the burden of an already overwhelmed healthcare system, as patients still need to see a clinician for a diagnosis. Moreover, many rely on black-box machine learning methods that identify patterns beyond human perception, making integration challenging because clinicians cannot readily verify such insights—an issue that contributes to algorithmic aversion \cite{Aristidou2022}. An open question is whether such systems can offer a more immediately practical application by supporting the patient–clinician interaction, making the elicitation of a comprehensive patient narrative both more efficient and comprehensive. This is particularly challenging in ADRD as symptoms are complex and often difficult for patients and their informants to describe, requiring probing questions and eliciting meaningful examples of their symptoms. Conversational AI agents, when integrated into the care pathway before a diagnostic visit, could serve as clinician-support tools to help elicit diagnostically relevant histories. This requires careful interaction and conversation design tailored to older adults—including question phrasing, turn-taking, and supportive feedback. These design choices shape how older adults perceive the agent \cite{pradhan2019phantomfriend, huang2025designingCAaging}: overly complex phrasing can increase cognitive load, while insufficient pauses or interruptions may discourage elaboration \cite{ rudnik2024carejournal, Kocaballi2022HealthCAReview, 10.1145/3719160.3736631}. Supportive cues and clarifications are not only functional aids but also signals of empathy that help participants feel understood and willing to share sensitive experiences \cite{ding2022talktive, xygkou2024mindtalker}. In this way, interaction design directly mediates the balance between clinical completeness and participant comfort, ensuring that the resulting interview produces narratives that are both actionable for clinicians and respectful of older adults’ communicative needs.

Building on these insights, we design voice-interactive conversational agents leveraging Large Language Models (LLMs), to collect a comprehensive patient narrative in collaboration with a major academic medical center's cognitive neurology clinic specializing in ADRD diagnosis and care. The agent conducts semi-structured interviews with patients and their informants to cover details that a panel of dementia specialists deemed important for the diagnosis and management of ADRD. The system incorporates interaction design elements for older adults with cognitive impairment~\footnote{We use the term cognitive impairment broadly to refer to clinically meaningful decline in one or more domains such as memory, attention, executive function, or language, exceeding age-expected changes. This usage encompasses mild cognitive impairment (MCI) as well as dementia.} such as allowing enough speaking time, avoiding interruptions, and providing interactional scaffolding (e.g., supportive cues, clarifications and structured follow-ups). In addition, the conversation design explicitly aims to help older adults with cognitive impairment articulate complex and often hard-to-describe experiences. We conduct a within-subject study with 30 older adults with suspected cognitive impairment caused by neurodegenerative disease, each assigned to both an agent-led and a clinician-led interview. We closely examine how older adults and their informants engage with the agent, and how interaction patterns influence the completeness of diagnostic information. These observations allow us to identify design considerations for older adult–oriented conversational agents and to highlight practical challenges for integrating such systems into clinical workflows. Through this work, we address the following research questions:
\begin{itemize}
    \item \textbf{RQ1:} How do older adults with cognitive impairment and their informants perceive and interact with agent-led interviews?
    \item \textbf{RQ2:} To what extent can a conversational agent elicit complex ADRD symptoms compared to dementia specialist interviews? 
    \item \textbf{RQ3:} What conversation design considerations are required to optimize user engagement and elicitation of complex symptoms with a conversational agent?
\end{itemize}

\noindent Note that for the second research question regarding symptom elicitation, we examine agreement between agent-led and clinician-led interviews to assess the feasibility of using conversational agents for systematic symptom elicitation. These comparisons were made to gather preliminary evidence about whether agents can capture clinically relevant and accurate information; they do not gather preliminary evidence of diagnostic performance. 

%% file: section/related_work.tex
\subsection{AI Technology for Older Adults}
Artificial intelligence (AI) technologies are being introduced into the lives of older adults, both to support everyday activities and to extend caregiving capacity. Socially assistive robots have been deployed to provide companionship, cognitive stimulation, and health reminders in aging populations \cite{davenport2019AIhealthcare, demiris2008, chan2009, peek2016, 10.1145/3643834.3661536} while also raising questions of ethics and acceptability \cite{abdi2018, fardeau2023, yen2024, karami2024}. Smart home systems and sensor‐based monitoring technologies have similarly been explored for supporting independent living, detecting falls, or tracking activities of daily living \cite{chan2009,peek2016}. In long‐term care contexts, voice assistants and other AI‐enabled devices have been appropriated for music, information seeking, and social connection \cite{Upadhyay2023LongTermVA,Grigorovich2024LTCVAs,Brewer2022OlderAdultsHealthVAs,abdollahi2022artificial}, showing both promise and challenges around accessibility, privacy, and sustained engagement \cite{kim2021longitudinalVA, nallam2020benefitsbarriers, 10.1145/3555089, 10.1145/3313831.3376529}. Beyond daily living support, AI systems are increasingly studied for applications in chronic illness care, depression detection, and other age-related health conditions \cite{feng2025ai_geriatric_chronic, fang2025_ai_physical_activity_mental_health, li2023ca_mental_health_meta}.

\subsection{Conversational AI for Cognitive Assessment and Dementia}
A growing body of work has explored the use of conversational agents (CAs) in health assessment and care, including applications for dementia \cite{10.1145/3706598.3713839, 10.1145/3659625, 10.1145/3731562}. For instance, TalkTive \cite{ding2022talktive} employs supportive listening behaviors (e.g., “Mm-hmm,” “Please, go on”) to engage older adults in neurocognitive screening based on pre-recorded audio, while CognoSpeak \cite{pahar2025cognospeak} deploys a mobile virtual agent to administer memory and fluency tasks, achieving high classification performance in detecting cognitive impairment. Other systems, such as entertainment‐based chatbots, engage users with news‐driven conversations and evaluate responses to identify potential cognitive impairment \cite{deArribaPerez2022entertainmentbot}. Speech biomarker approaches, integrated into robotic systems, analyze pronunciation, grammar disruption, and other features to generate composite indicators correlating with MMSE (Mini-Mental State Examination) scores \cite{perumandla2025robotCA}. Voice‐based versions of gold‐standard tests such as MoCA (Montreal Cognitive Assessment) have also shown strong performance in dialogue understanding \cite{pachecotorenzo2024digimoca}. Importantly, comparative evaluations suggest that conversational interaction with AI agents may impose less psychological burden than traditional clinician-led evaluations \cite{igarashi2024AIhumanstress}. However, few studies have directly compared agent-led interviews with clinician practices or examined how conversation design shapes patient narratives and the completeness of diagnostic information.

\subsection{Interaction Design for Older Adults in Conversational Systems}
Research on conversational systems for older adults examines how these technologies can align with their communication practices, values, and contexts. Older adults often personify or ontologically categorize voice assistants (VAs)—as a “phantom friend” or a “box with information”—with downstream implications for trust and expectations \cite{pradhan2019phantomfriend, huang2025designingCAaging}. Work in caregiving networks further explores limitations of off-the-shelf CAs (i.e., rigid cue–response patterns, weak personalization, breakdowns in multi-user talk) and calls for role-aware designs that acknowledge care responsibilities and accountability \cite{10.1145/3610170}. Field studies in long-term care communities show older adults integrating VAs into everyday routines (e.g., music, information lookups, reminders) and, at times, using them in shared spaces—raising considerations for shared devices and multiparty interaction \cite{10.1145/3571884.3597135, diederich2023semipublicVA}. Across this literature, recurring design levers include slower pacing and supportive prompts that allow longer turns, as well as robust repair when breakdowns occur. Beyond older adult–specific contexts, studies of conversational systems more broadly highlight strategies to recover from errors and maintain trust, for example, sincere apologies and clear ownership of mistakes \cite{10.1145/3491102.3517565}, and a range of system- and user-initiated repair strategies such as confirmation, rephrasing, and explicit clarification \cite{10.1145/3640794.3665558, zheng2022polyadic}.

%% file: section/method.tex
\subsection{System Overview}
\label{subsec:system_overview}
\begin{figure}[h]
  \centering
  \includegraphics[width=\linewidth]{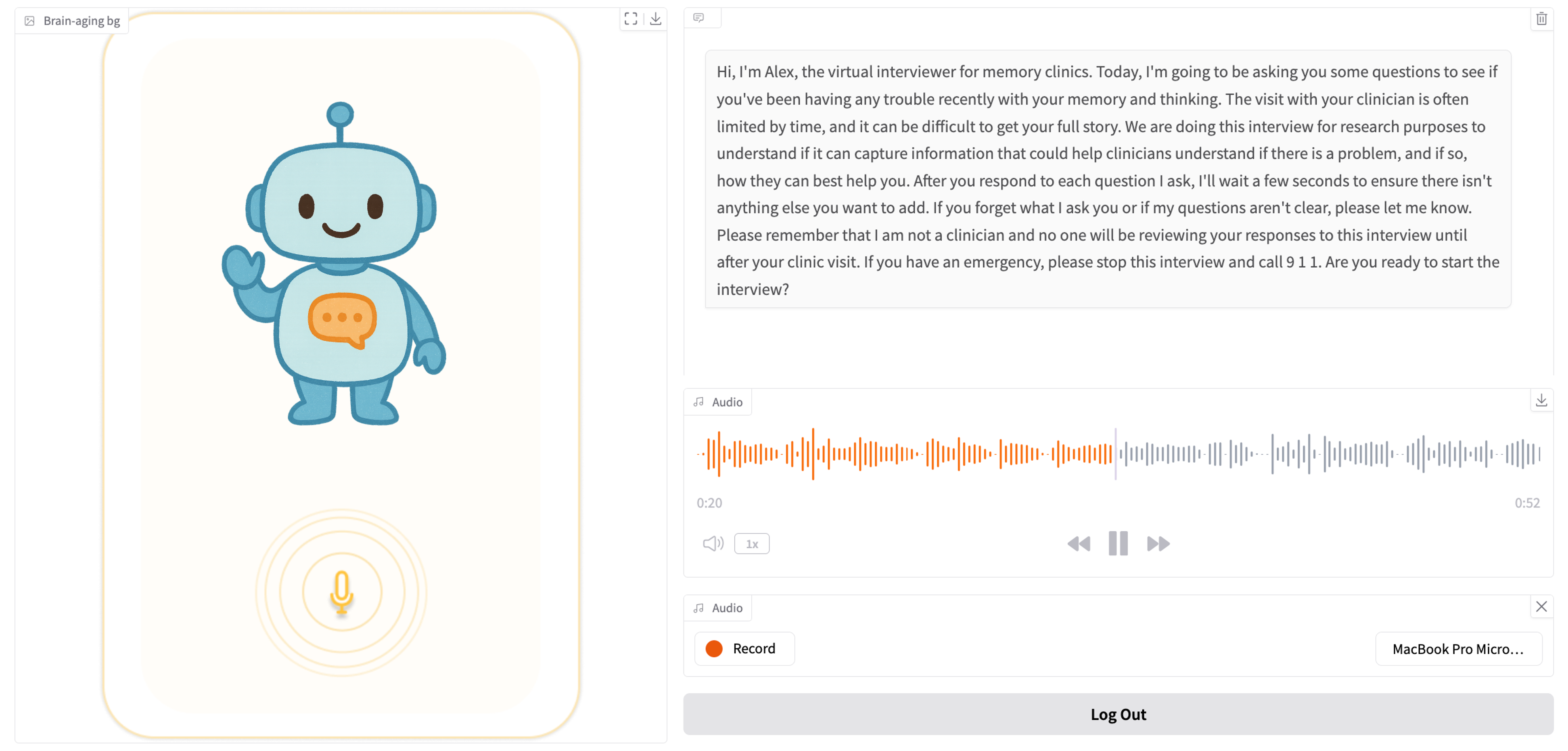}
  \caption{Web interface of the agent as seen by patients and informants during an interview. The left panel shows the animated agent avatar, while the right panel displays the real-time transcript, audio waveform, and playback controls.}
  \label{fig:system-ui}
\end{figure}

Our long-term aim is to deliver a phone call–based conversational agent that can be used remotely by older adults and their informants. However, in this early-stage pilot study, we developed a voice-interactive system that was deployed on a secure web platform and tested in-person with patients and their informants (e.g., spouse or adult child) before visits to a large urban cognitive neurology clinic. Figure~\ref{fig:system-ui} shows the interface presented to participants during the interview. The system design was guided by interaction principles for older adults and incorporated support for multi-party participation (see Section \ref{subsec:interaction_design}). The system conducts semi-structured diagnostic interviews using a specialist-developed question set for early ADRD diagnosis adapted from the Assessment of Cognitive Complaints Toolkit for AD (ACCT-AD). \cite{CADC2018manual, CADC2018toolkit}. This set contains approximately thirty topic areas, such as \textit{Memory Difficulties}, \textit{Language Difficulties}, \textit{Personality Changes} and \textit{Motor Changes}. Each topic area has one main question and three to four follow-up questions probing for specific symptoms or contextual details based on the participant’s responses. These conditional branching rules are embedded in the system prompt provided to the conversational agent. This structure enables systematic coverage of key diagnostic areas while supporting natural branching in the conversation. The entire system was deployed in a secure AWS environment and is powered by Claude 3.5, a large language model by Anthropic~\cite{anthropic2024claude35}, through the Bedrock API\footnote{Amazon Bedrock is a managed service that provides secure API access to foundation models, ensuring compliance with institutional security requirements.}. The web interface was developed using Gradio~\cite{gradio}. Audio is processed in real time using Whisper~\cite{radford2023whisper} for automatic speech recognition (STT) and Kokoro~\cite{kokoro82m_hf_2024} for text-to-speech (TTS). All audio recordings and corresponding text transcripts were stored with timestamps for post-hoc analysis, and post-interview experience feedback from participants was also collected for qualitative analysis using REDCap \footnote{REDCap (Research Electronic Data Capture) is a secure, web-based platform designed for building and managing online surveys and databases, widely used in academic and clinical research.}.  

\subsection{Interaction Design for Older Adults}
\label{subsec:interaction_design}
Interaction design was refined iteratively through observation. A study team member attended all agent-led interviews, monitoring participant reactions to different design choices from the side of the room without interfering. Insights from these real-time observations informed improvements throughout the study to question design (Section \ref{subsubsec:quation_design}), interview scaffolding (Section \ref{subsubsec:scaffolding}), turn-taking and latency control (Section \ref{subsubsec:latency_control}).

\subsubsection{Question Design}
\label{subsubsec:quation_design}
Initial questions consisted mainly of yes/no formats, with the only adaptation being a prompt for the agent to request examples of symptoms when participants responded "yes." This format elicited primarily short yes/no responses and only brief examples with the first five participants, at which point the question set was substantially revised to mimic a clinician interview style that begins with open-ended questions, then uses follow-up questions to explore participants' responses in greater depth. Due to these significant changes in the question set, the first five participants were not included in this analysis. 

\subsubsection{Interview Scaffolding}
\label{subsubsec:scaffolding}
The interview flow employed a two-part system prompt: high-level interview instructions (Figure~\ref{fig:header_prompt}) and topic-specific scripts (Figure~\ref{fig:prompt_reduced_empathy}). The high-level instructions defined the agent’s role, tone, and overall interaction guidelines, while the topic scripts ensured systematic coverage of each diagnostic area. The scaffolding strategy combined supportive cues, clarifications, and structured follow-ups to guide participants through each topic \cite{10.1145/3706598.3713999, sanjeewa2024empathic, 10.1145/3637320}. Supportive cues included short, concrete examples to aid memory recall (e.g., “in the past couple of years,” “while walking or driving”), phrased simply to enhance accessibility for older adults. When participants had difficulty understanding the question, the agent rephrased them in a concise manner to ensure understanding without overwhelming the participant. To prevent topic drift, the agent issued a single, polite redirection before proceeding to the next topic, keeping the session within a reasonable length. 

\begin{figure}[h!]
\small
\begin{tcolorbox}[width=\linewidth, colback=white, colframe=black, title=\textbf{\# Cognitive Decline Interview Instructions}]
Play the role of an empathetic, expert cognitive neurologist (also called behavioral neurologist). The user you speak to is an elderly patient that may or may not have dementia. Your goal is to interview them to gather information for a possible diagnosis and understand if their symptoms are hindering activities in their daily life (i.e., understanding their level of cognitive status). The patient may have someone with them (an informant) to help them answer: you should always speak directly to the patient, but answers from either the patient or the informant are acceptable. If the informant is not speaking, ensure they know we want to hear their answers as well. Do NOT allow the patient or informant to go off-topic, and tactfully redirect them towards the interview topics if they start talking about unrelated things. However, to ensure the interview is not too long, only try to redirect once for each question. Make sure to always be polite and allow the patient to take their time. Avoid repeating what the patient or informant says, and do not explain your thought process.\\

\vspace{-0.2cm}
> Below is a script you are using to guide your interview for the current topic. Please conduct the interview according to the content of this script only - this is very important to ensure the interview covers these questions in a timely manner.
\end{tcolorbox}
\vspace{-0.5cm}
\caption{High-level interview instructions provided to the conversational agent, outlining the agent’s role, tone, and interaction constraints for the interview with older adults and their informants.}
\vspace{-0.5cm}
\label{fig:header_prompt}
\end{figure}

\begin{figure}[h!]
\small
\begin{tcolorbox}[width=\linewidth, colback=white, colframe=black, title=\textbf{\#\# Topic Script: Memory Difficulties}]
\textbf{Main question:} Have you noticed any recent problems with your memory? \\
\textbf{\text{$\;$} > If yes:} Tell me more about that by providing an example. \\
\textbf{\text{$\:\;\;\;\;$} > Then:} Sometimes patients who describe difficulty with memory also describe problems with misplacing items or relying more on notes. Do any of these sound familiar? \\
\textbf{\text{$\:\;\;\;\;$} > Then:} How about trouble with remembering recent conversations, or asking repetitive questions? \\  
\textbf{\text{$\:\;\;\;\;$} > Then:} How long has this been going on? \\
\textbf{\text{$\:\;\;\;\;$} > Then:} Have these difficulties made doing some activities or tasks more difficult? \\
\textbf{\text{$\;$} > If no:} Just to make sure this question is understood, any difficulties misplacing items, relying on notes, trouble with recent memory, and asking repetitive questions? \\
\textbf{\text{$\:\;\;\;\;$} > If yes:} Tell me more about that by providing an example. \textbf{\text{$\:\;\;\;$} > If no:} Move on to next topic.
\end{tcolorbox}
\vspace{-0.5cm}
\caption{Example of a topic-specific script (\textit{Memory Difficulties}) containing a main question and conditional follow-ups, designed to elicit diagnostically relevant details.}
\label{fig:prompt_reduced_empathy}
\end{figure}
\vspace{-0.1cm}

\subsubsection{Turn-taking and Latency Control}
\label{subsubsec:latency_control}
Turn-taking was iteratively refined to address both the communication needs of older adults with suspected cognitive impairment and the multi-party nature of the sessions (patient and informant). The interface provided explicit visual indicators to signal when the agent is speaking and when it is actively listening. Unlike typical conversational agent settings that minimize response latency, longer latency was preferred here to avoid interrupting participants, whose slower speech pace and extended pauses often reflected effortful recall while formulating answers. We allowed up to 5 seconds of silence before the agent takes its turn, extending this threshold to 10 seconds for questions that required more elaborate descriptions. In multi-speaker settings, the agent addressed each role separately to maintain predictable alternation and reduce ambiguity about who should respond. In an effort to mimic a dementia specialist interview as much as possible, the patient and informant were interviewed together. For most clinician visits, it is not practical to interview the patient and informant separately. 

\subsection{Study Design and Participants}
\label{subsec:study_design}

\noindent \textbf{Study Overview.} $\;$ This study evaluated a voice-interactive conversational agent for ADRD diagnostic interviews through three complementary analyses. We conducted conversation analysis of all agent-led interviews (n=30) and collected user experience surveys (n=19) to examine how older adults perceive and interact with the agent (\textbf{RQ1}, Section~\ref{sec:user_experience}). To assess the agent's ability to elicit complex ADRD symptoms (\textbf{RQ2}), we compared agent-led and clinician-led interviews through analyses of sensitivity, specificity, ambiguity, and coverage (Section~\ref{sec:symptom_elicitation}). To identify optimal conversation design (\textbf{RQ3}), we systematically evaluated agent interaction behavior across two prompting strategies (Section~\ref{subsec:result_bot_interaction}). All participants completed agent-led interviews; a subset also had clinician interviews available for comparative analysis.

\vspace{0.2cm}
\noindent \textbf{Within-Subjects Comparison Design.} $\;$ For the comparative symptom elicitation analysis, we conducted a fixed-order, within-subjects study design in which participants completed an agent-led interview followed by a clinician-led interview either the same day or up to 3 months later. This fixed-order design reflects the intended deployment model where agents gather preliminary information before specialist consultation. We chose not to counterbalance order because the agent would always operate first in real-world deployment, and testing the reverse order could potentially result in artificially inflated performance that wouldn't translate to real-world practice. The first interviewer (agent) must work with patients and/or informants who are approaching symptom recall potentially for the first time, while the second interviewer (clinician) benefits from patients having already reflected on their symptoms. 

The variability in timing between agent interview and clinician interview also reflects real world practicalities. At the cognitive neurology clinic where this agent was studied, patients complete neuropsychological testing and intake questionnaires at an initial visit, followed by a specialist consultation the same day or within several months depending on clinician availability. The agent interview was integrated into this existing workflow, occurring at the time of neuropsychological testing. In the context of slowly progressive cognitive impairments typical of ADRD, clinically significant symptom progression over a 3-month period is unlikely except in rare cases of rapidly progressive dementia. This clinic's experience administering questionnaires using this same timing supports this approach.  

The agent followed the specialist-developed script (Section~\ref{subsubsec:scaffolding}), while the clinician was blinded to the agent interview so that they would conduct their interview using their usual clinical practice. All clinicians at the cognitive neurology clinic agreed to participate. This research was approved by the Institutional Review Board (IRB) and all participants provided informed consent.

\vspace{0.2cm}
\noindent \textbf{Data Security, Privacy, and Safety Monitoring.} $\;$ The conversational agent underwent institutional IT security review to ensure HIPAA compliance. The system and all data were hosted within a secure, HIPAA-compliant AWS account with encryption and role-based access controls. Audio recordings and transcripts were treated as Protected Health Information (PHI), as unique patient narratives could potentially identify individuals despite removal of standard identifiers. For participant safety, both the consent process and the conversational agent explicitly informed participants that transcripts would not receive immediate clinical review. A study team member reviewed all agent transcripts within 24 hours for safety concerns 
(e.g., homicidal or suicidal ideation). When consensus adjudication (Section~\ref{subsec:symptom_elicitation}) identified clinically relevant information in agent transcripts that was absent from clinician interviews, the reviewing neurologist contacted the treating clinician to ensure appropriate clinical follow-up.

\vspace{0.3cm}
\noindent \textbf{Participants.} $\;$ Adults with suspected ADRD were recruited from a large urban cognitive neurology clinic. Inclusion criteria only required sufficient hearing and speech ability to participate in spoken interviews, including participants who used functional hearing aids. Of the initial 35 participants who were interviewed by the conversational agent, data from first 5 were not included in the final analyses as explained above (Section~\ref{subsubsec:quation_design}). The next thirty participants (Table \ref{tab:participant}) were included in the analysis. Eighteen participants attended the session with an informant (typically a spouse or adult child) who could provide supplemental information. All participants who could not completely live independently due to cognitive difficulties had an informant present. 

\begin{table}[h!]
\centering
\begin{tabular}{lll}
\toprule
\multicolumn{1}{l}{\textbf{Demographic Factor}} & \textbf{} & \multicolumn{1}{l}{\textbf{Summary Statistics}} \\
\midrule
Gender & \small{Count (\%)} & \texttt{Female:} 20 (66.7\%),$\;\;$\texttt{Male:} 10 (33.3\%) \\
Race/Ethnicity & \small{Count (\%)} & \texttt{Asian:} 1 (3.3\%),$\;\;$\texttt{Black:} 6 (20.0\%),$\;\;$\texttt{White}: 23 (76.7\%) \\
Age & \small{Mean (SD) Range} & 72$\:\;$(9.8)$\:\;$47--89 \\
Years of Education & \small{Mean (SD) Range} & 15.6$\:\;$(1.9)$\:\;$12--20 \\
\bottomrule
\end{tabular}
\caption{Participant Demographic Information (n=30)}
\label{tab:participant}
\vspace{-0.3cm}
\end{table}

\begin{table}[h]
\centering
\begin{tabular}{p{4.5cm}cp{5.5cm}c}
\toprule
\textbf{Analysis} & \textbf{n} & \textbf{Exclusion Reason} & \textbf{Section} \\
\midrule
Agent conversation analysis & 30 & None & 4.1, 4.3 \\
\midrule
User experience surveys & 19 & 11 did not complete due to time constraints or non-response to email requests & 4.1 \\
\midrule
Symptom elicitation: Sensitivity/Specificity & 25 & 4 rescheduled without notifying team (no clinician recording); 1 had unblinded clinician& 4.2.2 \\
\midrule
Symptom elicitation: Ambiguity/Coverage & 23 & Additional 2 consultative visits excluded (abbreviated clinician assessments would inflate clinician rates)& 4.2.3 \\
\bottomrule
\end{tabular}
\caption{Sample composition across analyses}
\label{tab:sample_composition}
\end{table}

\vspace{-0.3cm}
\noindent \textbf{Procedure.} $\;$ Sessions were conducted in a private room at the clinic. In the agent-led condition, participants sat side-by-side with their informant in front of a laptop running the web-based voice interface (Figure~\ref{fig:system-ui}), with two microphones capturing speech from each speaker. Participants were assigned to one of two agents, each using a different prompting strategy (15 participants per condition):
\begin{itemize}
    \item \textit{Full-script prompting}, where the agent received the complete set of approximately 30 topic scripts in a single system prompt at the start of the session.
    \item \textit{Sequential prompting}, where the agent was given one topic script at a time, proceeding to the next only after the agent determined that the current topic had been fully addressed.
\end{itemize}

\noindent We included this experimental condition to examine whether the prompting method influences the agent’s instruction-following behavior, user interaction styles, and symptom elicitation.  All interviews were audio-recorded, transcribed, and checked for transcription accuracy.  After each interview, participants completed a short questionnaire rating their experience based on the validated Acceptability E-Scale (AES) \cite{Tariman2011AcceptabilityEScale, Philip2017VirtualHuman}.

\subsection{Evaluation Measures}
\label{subsec:eval}
We evaluated the system across three complementary dimensions—\textit{user experience}, \textit{symptom elicitation} and \textit{agent interaction behavior}. We also examined how these dimensions interact, for example, how agent prompting strategies influenced its interaction patterns, how these differences shaped participant engagement, and how both, in turn, affected the clinical quality of interview transcripts. Interviews that ended early, regardless of the reason, were included in all analyses.

\vspace{0.2cm} 
\noindent \textbf{User Experience.} $\;$ After each interview, participants completed a six-item questionnaire evaluating their experience with the agent interview. Each item used a 5-point Likert scale (5 = very positive, 1 = very negative) and assessed: (1) \textit{Ease of use}, (2) \textit{Understandability of questions}, (3) \textit{Enjoyment of the interaction}, (4) \textit{Helpfulness in describing symptoms}, (5) \textit{Acceptability of the time required}, and (6) \textit{Overall satisfaction}. The full wording of the six questionnaire items is shown in Figure~\ref{fig:patient_satisfaction}. Open-text responses captured qualitative feedback.

\vspace{0.2cm} 
\noindent \textbf{Symptom Elicitation.} \label{subsec:symptom_elicitation} $\;$ One cognitive neurologist~\footnote{A clinician with a subspecialty in neurodegenerative disease, board certified in neurology as well as certified by the United Council for Neurologic Subspecialties (UCNS) in behavioral neurology and neuropsychiatry} (AB) and one of two trained study team members (OA and AN) independently labeled each transcript from both agent-led and clinician-led interviews then met to resolve discrepancies through consensus. Each interview was labeled with 32 symptoms as \textit{present}, \textit{absent} (symptoms that are denied), \textit{ambiguous}, \textit{past history} (resolved or sufficiently treated), \textit{asked but unanswered}, or \textit{not discussed}. These 32 symptoms were identified by two cognitive neurologists (AB and HR) as comprehensive domains relevant to ADRD diagnosis and management. 

Inter-rater reliability prior to consensus adjudication was assessed using Cohen's kappa. A labeling guidebook was created and expanded as needed based on consensus discussions. For cases where agent and clinician transcripts showed discrepant labels of present, absent, or ambiguous for the same symptom, the same annotators independently reviewed both transcripts to establish a gold standard determination again through consensus adjudication. This cross-transcript review sometimes revealed that different interview approaches elicited complementary information. For example, in-depth questioning in one interview might reveal detailed descriptions that clearly indicated hallucinations, even though the patient denied having hallucinations in both interviews due to not recognizing their experiences as qualifying for this symptom. If the presence or absence of a symptom remained unclear even after reviewing both transcripts, the symptom was labeled as ambiguous in the gold standard.

For ambiguity and not-discussed rate analyses, we focused on a subset of 21 symptoms (Table~\ref{tab:symptoms}) that two cognitive neurologists (including the cognitive neurology division director) identified as both frequently elicited by clinicians at this clinic and clinically important for comprehensive evaluation. These represent symptoms that should ideally be assessed for every new dementia patient regardless of suspected diagnosis, accounting for expected practice variation while focusing on clinically meaningful domains. These covered symptoms that could uncover modifiable contributors, uncover common copathologies (e.g., Alzheimer's disease and Lewy Body disease are common copathologies), or identify commonly missed diagnoses. 

We then conducted this evaluation using four measures: 

\begin{itemize}
    \item (1) \textit{Sensitivity} — the proportion of symptoms present in the gold standard that were also correctly identified in the agent transcript.
    \item (2) \textit{Specificity} — the proportion of symptoms absent in the gold standard that were also marked absent in the agent transcript.
\end{itemize}

\noindent For these two metrics, the cross-transcript consensus determination served as the gold standard. The analysis for these two metrics was restricted to labels of "present" or "absent". We included 25 participants with available clinician transcripts, including two consultative visits where patients presented with established diagnoses from outside clinics. The symptoms labeled as present or absent meant they dove into these symptoms with sufficient detail.

\begin{itemize}
    \item (3) \textit{Not Discussed rate} — the proportion of the 21 clinically important symptoms omitted entirely from each interview (agent or clinician).
    \item (4) \textit{Ambiguity rate} — the proportion of the 21 clinically important symptoms discussed but for which presence/absence could not be clearly determined. 
\end{itemize}

\noindent For the latter two measures, we excluded the two consultative visits (n=23), as the abbreviated symptom exploration in cases with established diagnoses would inflate not-discussed rates and potentially bias ambiguity assessments. In consultative visits, clinicians may not pursue clarification of ambiguous symptoms when confident in an established diagnosis by an outside clinic. 

For all four measures, we calculated both overall proportions and symptom-specific rates. While overall proportions provide summary measures of agent performance, symptom-specific analysis ensures comprehensive evaluation across all symptom domains. This approach prevents masking of poor performance in specific symptom areas that could be obscured by overall proportions, ensuring that both systematic gaps (high not-discussed rates) and persistent ambiguity (high ambiguity rates) are detected at the individual symptom level.  

\begin{table}[h]
\small
\centering
\begin{tabular}{lll}
\toprule
\textbf{Cognitive/Behavioral} & \textbf{Functional/Motor} & \textbf{Sleep/Psychiatric} \\
\midrule
Memory problems & Basic ADL & Sleep disturbances \\
Language difficulties & Instrumental ADL & Sleep apnea \\
Executive dysfunction & Falls & REM sleep behavior disorder \\
Visuospatial difficulties & Description/cause of fall & Depression \\
Hallucinations & Gait changes & Anxiety \\
Delusions & Balance & Irritability \\
Socially inappropriate behaviors & Tremor & \\
Family history & & \\
\bottomrule
\end{tabular}
\vspace{0.2cm}
\caption{Core ADRD symptoms selected for ambiguity and not-discussed rate analyses 
(n=21). These symptoms were identified through clinical consensus as both frequently 
elicited in routine practice and important for comprehensive dementia evaluation 
regardless of suspected diagnosis.}
\label{tab:symptoms}
\vspace{-0.3cm}
\end{table}

\noindent This pilot study was exploratory without prespecified primary or secondary endpoints and therefore results are treated as hypothesis generating. For ambiguity and not discussed rate analyses, we treated agent and clinician interviews as paired observations. Because the agent attempts all symptoms while clinicians selectively elicit those they judge relevant, direct comparison of not-discussed rates is inherently asymmetric; we therefore report not-discussed rates descriptively. For ambiguity rates, which were restricted to symptoms actually discussed, we compared agent and clinician values using Wilcoxon signed-rank tests across symptoms. Overall proportion differences were assessed using exact binomial tests. When stratifying by agent prompting strategy (full-script vs. sequential), we applied Mann-Whitney U tests for between-group comparisons for both rates. Significance was set at p < 0.05 with two-sided tests throughout. For sensitivity and specificity analyses, overall metrics included all available data, while per-symptom analyses were limited to symptoms with adequate sample sizes of at least five cases. Confidence intervals for sensitivity and specificity were calculated using Wilson score intervals. 

\vspace{0.2cm} 
\noindent \textbf{Agent Interaction Behavior.} $\;$ We analyzed the agent-led transcripts to assess whether the agent covered all required topic scripts, whether its utterances were appropriate and safe, and how it adapted its questions and follow-ups. We developed a detailed annotation guideline for evaluating agent utterance quality, informed by prior work on conversational quality assessment \cite{finch-etal-2023-dont}. The guideline covers five dimensions: (1) \textit{Topic script coverage}, and (2) \textit{Topic script alignment}, (3) \textit{Response politeness}, (4) \textit{Correct understanding of patient/informant input} and (5) \textit{Provision of opportunities for participants to respond}. The full annotation guideline is provided in \ref{appendix:annotation_guideline}. Given the scale of the dataset, we used an LLM-based annotation approach \cite{chiang-lee-2023-large} to label all agent utterances, as a pragmatic aid to support large-scale analysis. Following the guideline, we automatically labeled all agent utterances, using Claude 4 Sonnet \cite{claude4}. To assess the reliability of these automated annotations, two trained human annotators independently annotated a randomly sampled set of five interview transcripts (\texttt{P9}, \texttt{P33}, \texttt{P15}, \texttt{P28}, \texttt{P31}). Each interview contains about 50--60 agent utterances (Table~\ref{tab:utterance_count}), resulting in several hundred agent utterances used for agreement analysis. Human annotations were then compared against the LLM labels to evaluate alignment and to contextualize the reliability of the automated coding. We computed inter-rater agreement using both raw agreement (i.e., the proportion of identical labels) and Krippendorff’s $\alpha$ \cite{krippendorff2011computing}, which is a chance-corrected measure, between each human annotator and the LLM as well as between the two human annotators. Agreement was calculated separately for each annotation dimension.

%% file: section/result.tex
\subsection{User Experience}
\label{sec:user_experience}
\begin{figure}[h]
  \centering
  \includegraphics[width=\linewidth]{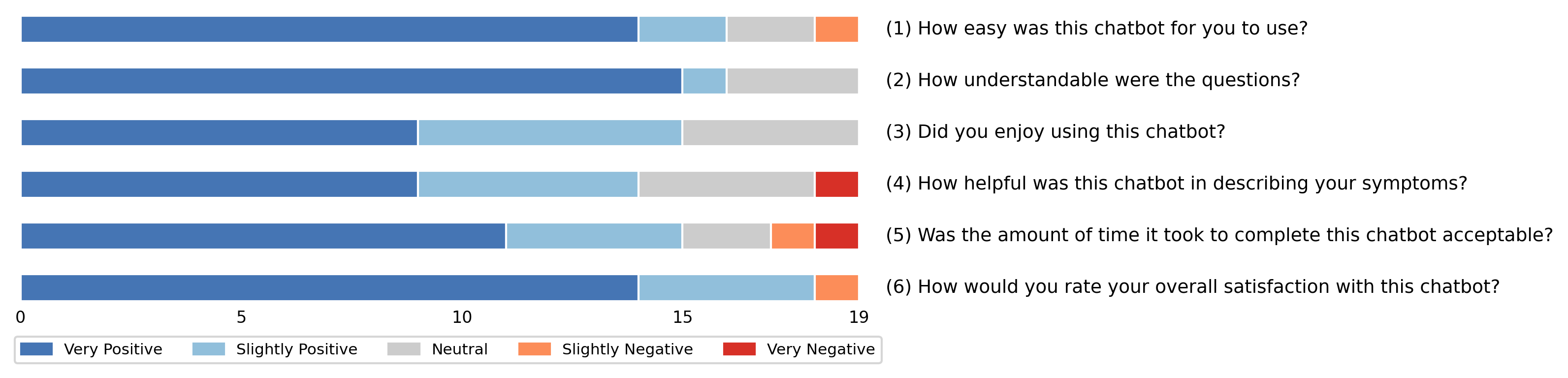}
  \caption{Patient satisfaction ratings (n=19) across six questionnaire items, each measured on a 5-point Likert scale. (\textcolor{blue}{5} = very positive, \textcolor{red}{1} = very negative)}
  \label{fig:patient_satisfaction}
\end{figure}

As shown in Figure \ref{fig:patient_satisfaction}, participants’ responses were consistently positive across all items. However, item (4) \textit{Helpfulness in describing symptoms} and (5) \textit{Acceptability of the time required} elicited a small number of strongly negative responses, typically linked to technical or conversational breakdowns. Of the 30 participants, 19 completed the questionnaire. 

\noindent Open‐text responses provided qualitative insights into participants’ experiences. Participants described the agent as \textit{“actually quite fun”} (\texttt{P14}), \textit{“enjoyable”} (\texttt{P9}, \texttt{P12}), and \textit{“entertaining to use”} (\texttt{P13}). Some appreciated its non‐interruptive style and the opportunity for detailed sharing, for example:
\begin{quote}
\small “I appreciate she doesn’t interrupt and gave us time to talk.” (\texttt{P24})
\end{quote}
\begin{quote}
\small “I thought that the follow up questions worked well to get a better understanding of my issues.” (\texttt{P27})
\end{quote}
\begin{quote}
\small “It was nice not having the pressure of answering questions with a person.” (\texttt{P25})
\end{quote}
Others highlighted its potential clinical value. One participant noted its usefulness as a supplementary tool in contexts where long wait times make it difficult to see a doctor:
\begin{quote}
\small “She can be really useful to people who are waiting a long time to see a doctor.” (\texttt{P6})
\end{quote}
Another emphasized the agent’s thoroughness in conducting the clinical interview:
\begin{quote}
\small “I found this to be very thorough and felt it understood my information more than most doctors. It was affirming and accurate with the majority of my patient history. It was patient and took time to listen.” (\texttt{P7})
\end{quote}
\noindent Not all feedback was positive, however. Some participants reported issues with question design, such as \textit{“sometimes questions are jumbled”} (\texttt{P15}) and \textit{“some repetitive questions”} (\texttt{P16}). One participant who responded negatively to the question (5) \textit{Was the amount of time it took to complete this chatbot acceptable?} (see Figure \ref{fig:patient_satisfaction}) remarked, \textit{“some of the questions should be broken down as opposed to 2 or 3 questions”} (\texttt{P23}). Some participants also noted clarity issues, such as being asked about concepts without sufficient explanation:
\begin{quote}
\small “Some questions asked about things without a definition, such as activities of daily living.” (\texttt{P19})
\end{quote}
\noindent In addition to conversation design issues, one participant reported cases where the system appeared to stall, underscoring the need for further backend process optimization:
\begin{quote}
\small “It locked up twice and was pretty slow a couple of other times. I tried my patience when it was working well. A faster processor might help.” (\texttt{P27})
\end{quote}
\noindent In addition to patient-reported feedback, we analyzed the conversational dynamics to better understand how patients and informants engaged with the agent. We observed that patients initially spoke more slowly and deliberately with the agent compared to natural conversation, a pattern that diminished as interviews progressed. Table \ref{tab:utterance_count} summarizes participants' utterance counts comparing clinician-led interviews (n=25) and the two types of agent-led interviews (n=15 each). In the interviewer role, clinicians spoke most frequently (about 118 turns on average), whereas the agents produced fewer utterances (roughly 50–65 turns). Patients, in turn, accounted for a greater share of turns and contributed longer responses in the agent-led sessions, especially under the sequential prompting setting. Informants, while also producing longer responses in the agent-led setting, nonetheless accounted for a smaller share of turns compared to clinician interviews, shifting the patient-to-informant ratio from roughly 1.5:1 to nearly 3:1. In clinician-led interviews, interviewers averaged 118.3 utterances, with patients contributing 81.6 and informants 51.6. By contrast, in the agent-led setting (sequential), interviewers produced 66.1 turns, patients 57.5, and informants only 19.9. It reflects participants’ survey findings, which characterized the agent as patient and non-interruptive, and as easing the interpersonal expectations of a clinician interview.

\begin{table}[h]
\centering
\small
\resizebox{\linewidth}{!}{%
\begin{tabular}{c|ll|ll|ll}
\toprule
\textbf{} & \multicolumn{2}{c|}{\textbf{Interviewer}} & \multicolumn{2}{c|}{\textbf{Patient}} & \multicolumn{2}{c}{\textbf{Informant}} \\ \midrule
\textbf{Interviewer} & \multicolumn{1}{c}{\small{Count}} & \multicolumn{1}{c|}{\small{Length}} & \multicolumn{1}{c}{\small{Count}} & \multicolumn{1}{c|}{\small{Length}} & \multicolumn{1}{c}{\small{Count}} & \multicolumn{1}{c}{\small{Length}} \\ \midrule
Clinician & 118.3 (50.0 -- 188.0) & 31.1 (18.7 -- 48.3) & 81.6 (2.0 -- 155.0) & 13.1 (1.0 -- 28.0) & 51.6 (11.0 -- 105.0) & 19.9 (9.8 -- 61.2) \\
Agent \small{(Full-script)} & 49.3 (36.0 -- 65.0) & 30.9 (22.5 -- 55.5) & 47.4 (35.0 -- 61.0) & 31.8 (3.1 -- 127.5) & 17.0 (3.0 -- 55.0) & 36.6 (7.1 -- 51.5) \\
Agent \small{(Sequential)} & 66.1 (31.0 -- 107.0) & 32.9 (22.3 -- 48.0) & 57.5 (7.0 -- 105.0) & 39.5 (10.1 -- 130.5) & 19.9 (5.0 -- 74.0) & 45.3 (15.8 -- 101.6) \\ \bottomrule
\end{tabular}%
}
\vspace{0.2cm}
\caption{Participants’ utterance counts and average length per utterance (in words; word counts computed using whitespace delimitation), with means and ranges, across clinician- and agent-led interviews, separated by role (interviewer, patient, informant).}
\vspace{-0.7cm}
\label{tab:utterance_count}
\end{table}

\noindent Eleven of the 30 agent interviews ended prematurely, two of which were terminated by patients due to time burden (each over an hour long). Seven interviews were terminated by the chatbot due to misinterpretation of conversational content as a request to end the session. Two interviews were terminated by research staff due to time constraints (patients needed to leave for their clinician appointment).

\vspace{0.2cm}
\noindent \textbf{Safety Monitoring.} $\;$ Review of all agent interview transcripts within 24 hours identified no concerning content requiring immediate clinical intervention, including no reports of suicidal ideation or acute medical emergencies.

\subsection{Symptom elicitation}
\label{sec:symptom_elicitation}
\subsubsection{Inter-rater Reliability}
Agreement between annotators before consensus was substantial (Cohen’s $k$ = 0.72–0.73), supporting reliability of symptom labeling.

\subsubsection{Sensitivity and Specificity}
\label{subsec:sensitivity and specificity}
\begin{figure}[h]
  \centering
  \includegraphics[width=\linewidth]{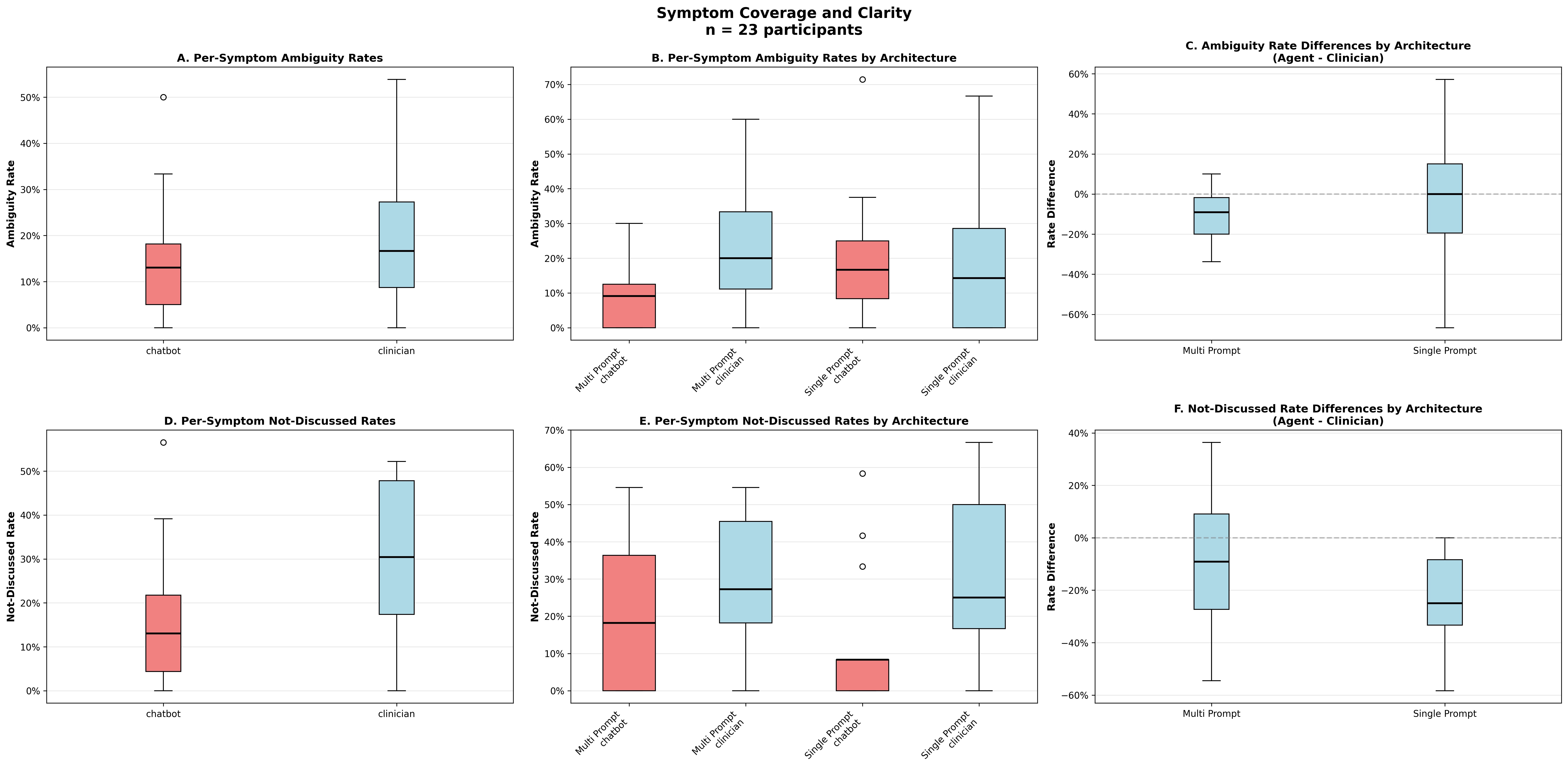}
  \vspace{-0.7cm}
  \caption{Box plots comparing per-symptom ambiguity rates (top row) and not-discussed rates (bottom row) between chatbot and clinician interviews (n=23). Left column (A, D) shows overall distributions across all architecture-assigned participants. Middle column (B, E) shows rates stratified by chatbot architecture type (Single Prompt vs Multi Prompt). Right column (C, F) shows the distribution of rate differences (Agent - Clinician) by architecture type. Box plots display median, quartiles, and outliers for per-symptom rates. Positive differences indicate higher rates for the agent compared to clinician interviews.}
  \label{fig:diagnostic}
\vspace{-0.1cm}
\end{figure}

Table \ref{tab:agent_vs_clinician_label_count} displays the count of symptom labels for the 25 paired agent and clinician interviews, and Table \ref{tab:sensitivity_specificity} and Figure \ref{fig:diagnostic} summarizes key measures of symptom elicitation from the agent-led interview transcripts. Across 25 paired agent-clinician transcripts, agent-led interviews achieved an overall sensitivity of 83.5\% and specificity of 100\% for symptoms labeled as clearly present or absent in the consensus gold standard. Sequential prompting (n=11) achieved higher sensitivity than full-script prompting (n=12) (87.6\% vs 78.7\%) with identical specificity. However, per symptom analyses were limited by small sample sizes. Among 18 symptoms with at least five cases that were present or absent in both transcripts, sensitivity ranged from 40-100\% and most estimates had confidence interval widths exceeding 40\%.

\begin{table}[h]
\centering
\begin{tabular}{lrr}
\toprule
\multicolumn{1}{c}{\textbf{Label}} & 
\multicolumn{1}{c}{\textbf{Agent Interview}} & 
\multicolumn{1}{c}{\textbf{Clinician Interview}} \\
\midrule
PRESENT          & 208& 171\\
ABSENT           & 371& 134\\
AMBIGUOUS        & 83& 73  \\
ASKED/UNANSWERED & 10  & 1   \\
PAST HISTORY     & 11  & 18  \\
NOT DISCUSSED    & 192& 478\\
\bottomrule
\end{tabular}
\caption{Distribution of Agent vs. Clinician Interview symptom labels}
\vspace{-0.3cm}
\label{tab:agent_vs_clinician_label_count}
\end{table}

\begin{table}[h]
\centering
\begin{tabular}{lrrr}
\toprule
\multicolumn{1}{c}{\textbf{Metric}} & \multicolumn{1}{c}{\textbf{Overall}} & \multicolumn{1}{c}{\textbf{Sequential Prompting}} & \multicolumn{1}{c}{\textbf{Full-Script Prompting}} \\
\midrule
Sensitivity (95\% CI) & 83.5\% (76.3--88.7)& 87.6\% (78.1--93.3)& 78.7\% (67.5--86.9)\\
Specificity (95\% CI) & 100\% (95.8--100)& 100\% (89.8--100)& 100\% (93.3--100)\\
\bottomrule
\end{tabular}
\caption{Sensitivity and specificity across all symptoms}
\vspace{-0.3cm}
\label{tab:sensitivity_specificity}
\end{table}

\subsubsection{Ambiguity and Not-Discussed Rates}
\label{subsec:ambiguity}
Analysis of the 21 clinically important symptoms (Table \ref{tab:symptoms}) revealed that agent interviews had a median per-symptom not-discussed rate of 13.0\% versus 30.4\% for clinicians. Ambiguity rates were comparable between interview modalities (median rates: 13.0\% agents vs 16.7\% clinicians; Wilcoxon signed-rank test, p = 0.19), with 47.9\% of ambiguous responses from agent interviews (exact binomial test, p = 0.71). Sequential prompting showed a trend toward lower ambiguity rates than full-script prompting (9.1\% vs 16.7\%; Mann-Whitney U test, p = 0.064), while not-discussed rates had no significant difference between prompting strategies (18.2\% vs 8.3\%; p = 0.165). Plots of ambiguity and not discussed rates can be seen in Figure~\ref{fig:diagnostic}.

\subsection{Agent Interaction Behavior}
\label{subsec:result_bot_interaction}
\begin{figure}[h]
  \centering
  \includegraphics[width=\linewidth]{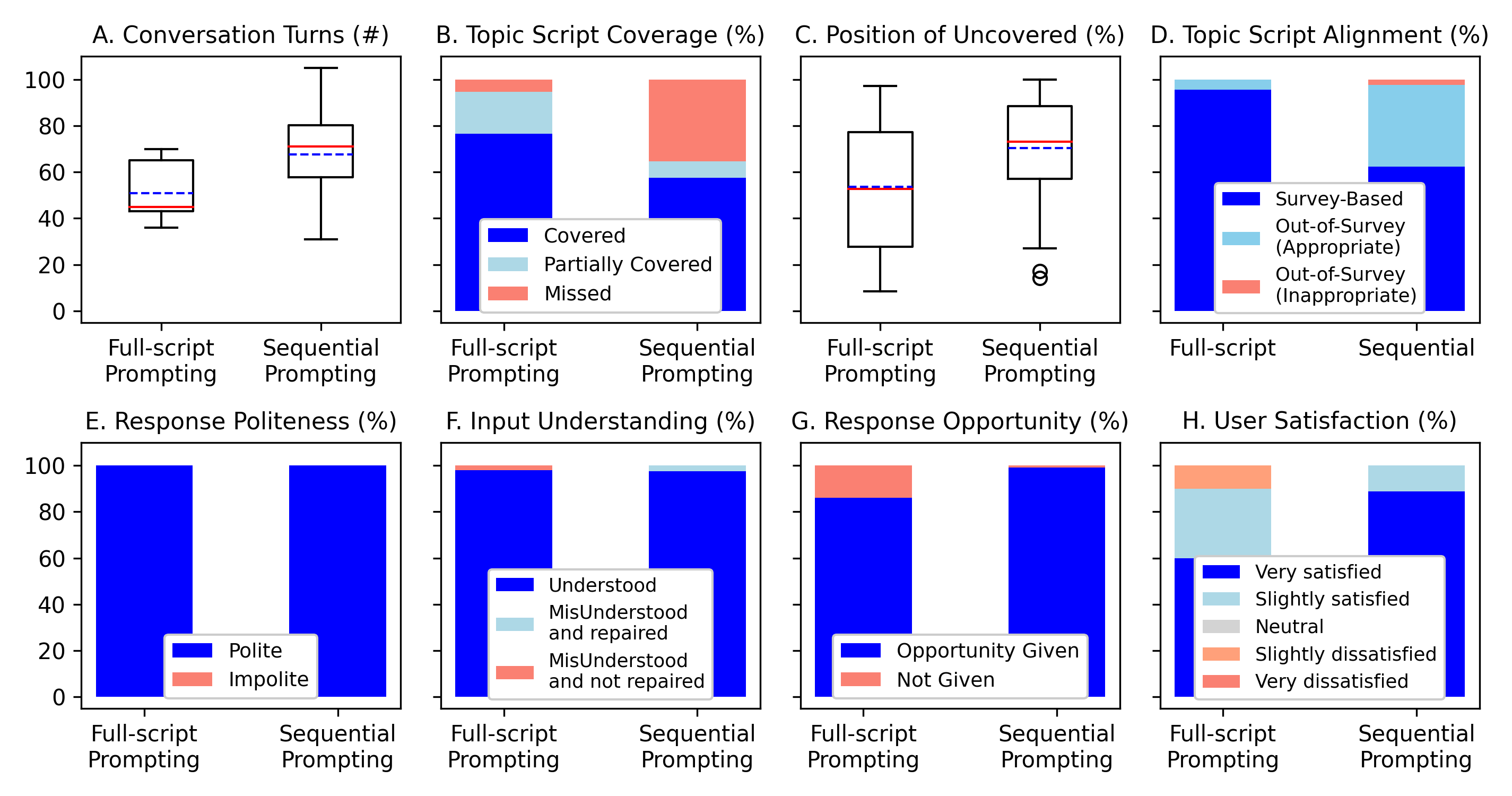}
  \vspace{-0.7cm}
  \caption{Utterance-level analysis of agent interaction behavior under the two prompting strategies. \textbf{(A) Conversation Turns} shows the conversation length, quantified by the total number of turn exchanges. \textbf{(B) Topic Script Coverage} indicates the proportion of topics \textit{covered}, \textit{partially covered}, or \textit{missed}. \textbf{(C) Position of Uncovered} indicates the normalized position of \textit{missed} or \textit{partially covered} in the sequence of topic scripts, where smaller values indicate earlier failures and larger values indicate later failures. \textbf{(D) Topic Script Alignment} shows whether utterances were directly \textit{survey-based}, \textit{out-of-survey but appropriate}, or \textit{out-of-survey and inappropriate}. \textbf{(E) Response Politeness}, \textbf{(F) Input Understanding}, and \textbf{(G) Response Opportunity} summarize utterance-level judgments of politeness, comprehension, and whether participants were given space to respond. \textbf{(H) User Satisfaction} compares overall satisfaction with the agent interaction across prompting strategies, corresponding to the sixth questionnaire item in Figure~\ref{fig:patient_satisfaction}.}
  \label{fig:bot_behavior}
\end{figure}

We present an utterance-level analysis of agent interaction behavior in Figure~\ref{fig:bot_behavior}. It compares two prompting strategies: \textbf{full-script prompting} (n=15), where the agent received all topic scripts at once, and \textbf{sequential prompting} (n=15), where it was given one topic script at a time and instructed to output the phrase \textit{move on to next topic} to trigger the next script. The analysis was derived from automatic LLM annotations, with reliability evaluated against human annotator judgments in Section \ref{subsec:human_annotation}.

\vspace{0.2cm} 
\noindent \textbf{Conversational richness and coverage trade-offs.} $\;$ 
Compared to full-script prompting, sequential prompting produced more conversation turns (A) and showed greater flexibility in topic script alignment (B), frequently generating appropriate out-of-survey questions (D). Consequently, sequential prompting promoted longer, more interactive conversations, frequently supplemented by clarifying probes that elicited further clinical detail. For instance, when a participant’s answer to \textit{“How long has this been going on?”} did not provide enough detail, the agent could adapt with a clarifying follow-up such as:
\begin{quote}
\small 
“Could you clarify—how long have you been noticing these changes in your movement speed, like walking more slowly or taking longer with buttons?” (agent utterance, sequential prompting, interview with \texttt{P19})
\end{quote}
This adaptivity contributed to higher sensitivity as shown earlier in Figure~\ref{fig:diagnostic}. However, the sequential design also introduced a distinct vulnerability: because topic transitions depended on the agent correctly producing the explicit \textit{move on to next topic} signal, failures to do so caused the agent to become stuck within a topic, while overproducing the signal caused it to skip multiple topics. These transition errors increased the likelihood of missed or only partially covered items, particularly in later sections of the script (C), thereby reducing overall topic script coverage. By contrast, full-script prompting tended to cover a wider range of topics overall, often addressing them at least partially rather than missing them entirely, though it offered little flexibility beyond the scripted survey questions. 

These differences were also reflected in interview duration. Duration data were available for 22 paired agent–clinician interviews. Agent-led interviews were longer overall (median 46 minutes, IQR 38–63) than the history-taking portion of clinician visits (median 25 minutes, IQR 17–30). Sequential prompting (n = 10) resulted in the longest interviews (median 62 minutes, IQR 55–81), compared with 25 minutes (IQR 17–32) for the corresponding clinician encounters, whereas full-script prompting (n = 12) produced shorter agent interviews (median 39 minutes, IQR 21–43) relative to clinicians (median 25 minutes, IQR 18–27).

\vspace{0.2cm} 
\noindent \textbf{Politeness and rapport.} $\;$ Across both strategies, all utterances were consistently rated as polite (E), with no antisocial or inappropriate behavior observed. Cases of misunderstanding user input (F) were relatively rare in both strategies; however, when they did occur, full-script prompting tended to proceed with the interview as scripted without repair, whereas sequential prompting more often flexibly repaired the misunderstanding before moving on. For example, when a participant had already described driving experiences and the agent later asked whether they drove, the participant responded that this had already been discussed, and the agent repaired the misunderstanding by saying:
\begin{quote}
\small 
“You're right. We already covered driving earlier when you mentioned the recent incidents including the tire incident on the expressway and the post at the gas station [...]” (agent utterance, sequential prompting, interview with \texttt{P28})
\end{quote}

\vspace{0.2cm} 
\noindent \textbf{Utterance design and response opportunities.} $\;$ Another important difference lay in how agent utterances were structured. With full-script prompting, the agent often attempted to compress multiple questions into a single turn—even when they belonged to different topics. This reduced the natural flow of the conversation and limited opportunities for participants to respond (G) fully to each question. For example, in one conversation the agent combined two distinct topics within a single utterance—first asking about changes in empathy and then shifting abruptly to a question about scams:
\begin{quote}
\small “That's great. We're asking about new changes in empathy or emotional reactions. Is your answer still no? Compared to the past, are you more open to scams or solicitations?” (agent utterance, full-script prompting, interview with \texttt{P31})
\end{quote}
In contrast, sequential prompting delivered one topic script at a time, which the agent was required to complete before moving on. This design promoted faithful execution of each topic, yielding cleaner turn-taking with one question per utterance and clear response opportunities for participants, resulting in a more conversational flow.

\subsubsection{Comparison with Human Annotation}
\label{subsec:human_annotation}
Table~\ref{tab:agreement} reports inter-rater agreement between each human annotator and the LLM, as well as between the two human annotators, computed on a randomly sampled set of five interviews. Overall, raw agreement is high across dimensions, while chance-corrected agreement varies by category, indicating that the LLM’s annotations are broadly aligned with human judgments at the utterance level. Across dimensions, Krippendorff’s alpha ($\alpha$) between the LLM and human annotators is generally comparable to human--human agreement, suggesting that automated labels fall within the range of variability observed between human annotators. Lower $\alpha$ values appear for several dimensions involving rare judgments (e.g., \textit{Out-of-Survey (b)}, \textit{Misunderstanding}, and \textit{Failure to Repair Misunderstanding}), despite high raw agreement. This pattern has been noted in prior work, which observes that chance-corrected agreement measures such as Krippendorff’s $\alpha$ tend to be conservative under low-prevalence or highly imbalanced label distributions \cite{krippendorff2011computing, artstein2008inter}. \textit{Survey Coverage} shows moderate agreement, with both raw agreement and $\alpha$ values lower than those observed for binary dimensions. This likely reflects the increased difficulty of the task due to its three-way categorical labeling scheme, which requires finer-grained distinctions among coverage states. Notably, human--LLM agreement on this dimension remains comparable to human--human agreement, suggesting that the LLM captures distinctions that also pose challenges for human annotators. For \textit{Antisocial / Impolite Responses}, all annotators consistently labeled all instances as non-antisocial, resulting in perfect raw agreement and undefined $\alpha$ due to the absence of label variance. We note that this human-annotated subset represents a limited sample of the full dataset, and the agreement analysis is intended to assess plausibility rather than to serve as a definitive validation of the automated annotations.

\begin{table}[h]
\centering
\small
\begin{tabular}{l|ll|ll|ll}
\toprule
                                   & \multicolumn{2}{c|}{\textbf{Human 1 --- LLM}}              & \multicolumn{2}{c|}{\textbf{Human 2 --- LLM}}              & \multicolumn{2}{c}{\textbf{Human 1 --- Human 2}}          \\ \cline{2-7} 
\textbf{Dimension} &
\multicolumn{1}{c|}{\makebox[1.2cm]{raw}} &
\multicolumn{1}{c|}{\makebox[1.2cm]{$\alpha$}} &
\multicolumn{1}{c|}{\makebox[1.2cm]{raw}} &
\multicolumn{1}{c|}{\makebox[1.2cm]{$\alpha$}} &
\multicolumn{1}{c|}{\makebox[1.2cm]{raw}} &
\multicolumn{1}{c}{\makebox[1.2cm]{$\alpha$}} \\ \midrule
Survey Coverage                    & \multicolumn{1}{r|}{70.3}  & \multicolumn{1}{r|}{61.4}                          & \multicolumn{1}{r|}{78.6}  & \multicolumn{1}{r|}{74.9}                          & \multicolumn{1}{r|}{76.0} & \multicolumn{1}{r}{64.2}                         \\
Out-of-Survey (a)                  & \multicolumn{1}{r|}{92.1}  & \multicolumn{1}{r|}{63.6}                          & \multicolumn{1}{r|}{92.6}  & \multicolumn{1}{r|}{62.9}                          & \multicolumn{1}{r|}{94.7} & \multicolumn{1}{r}{74.1}                         \\
Out-of-Survey (b)                  & \multicolumn{1}{r|}{90.1}  & \multicolumn{1}{r|}{58.8}                          & \multicolumn{1}{r|}{90.3}  & \multicolumn{1}{r|}{57.0}                          & \multicolumn{1}{r|}{93.4} & \multicolumn{1}{r}{64.7}                         \\
Antisocial / Impolite Responses    & \multicolumn{1}{r|}{100.0} & \multicolumn{1}{r|}{N/A}                           & \multicolumn{1}{r|}{100.0} & \multicolumn{1}{r|}{N/A}                           & \multicolumn{1}{r|}{100.0}  & \multicolumn{1}{r}{N/A}                          \\
Misunderstanding                   & \multicolumn{1}{r|}{94.1}  & \multicolumn{1}{r|}{64.8}                          & \multicolumn{1}{r|}{94.9}  & \multicolumn{1}{r|}{49.0}                          & \multicolumn{1}{r|}{95.5} & \multicolumn{1}{r}{52.0}                         \\
Failure to Repair Misunderstanding & \multicolumn{1}{r|}{90.5}  & \multicolumn{1}{r|}{63.6}                          & \multicolumn{1}{r|}{91.3}  & \multicolumn{1}{r|}{47.3}                          & \multicolumn{1}{r|}{95.5} & \multicolumn{1}{r}{52.2}                         \\
No Opportunity to Answer           & \multicolumn{1}{r|}{96.8}  & \multicolumn{1}{r|}{85.6}                          & \multicolumn{1}{r|}{97.3}  & \multicolumn{1}{r|}{71.6}                          & \multicolumn{1}{r|}{96.1} & \multicolumn{1}{r}{71.7}                         \\ \bottomrule
\end{tabular}
\vspace{0.1cm}
\caption{Inter-rater agreement between each human annotator and the LLM as well as between the two human annotators.}
\vspace{-0.3cm}
\label{tab:agreement}
\end{table}

%% file: section/discussion.tex
\subsection{Socio-technical Considerations in Designing Conversational Agents for Early Diagnosis of ADRD}
Our study shows that conversational agents for support of early ADRD diagnosis need to be designed as socio-technical systems, meaning that their effectiveness depends not only on reliable language processing but also on how they fit the communication styles of older adults, particularly those with cognitive impairment, and the dynamics of multi-party interaction. Prior research has found that older adults often treat voice assistants in different ways—e.g., as friendly companions, as simple appliances, or as fragile listeners—and these perceptions shape how they talk to the system \cite{pradhan2019phantomfriend, huang2025designingCAaging}. In our study, many participants began by speaking slowly and over-articulating, which can be understood as a way of helping the agent “hear better.” This behavior can be understood as a user-led repair strategy \cite{10.1145/3641026}, anticipating possible recognition failures. Prior work has shown that such adjustments are common when people interact with conversational agents, as users actively compensate for expected system limitations \cite{luger2016like, porcheron2018voice, 10.1145/3449101}. Such behaviors indicate that older adults are not merely passive recipients but active participants in sustaining the interaction \cite{salome2025codesignCAolder}. System design can build on this by explicitly scaffolding such strategies, for example by providing clear listening indicators that reassure users their effort is being understood \cite{10.1145/3555089}. In our study, we observed that participant behavior seemed to change the most when the agent gave concise summaries of what the participant had stated and/or attempted to clarify participant responses, resulting in the participant then speaking in a more natural and conversational manner. 

In addition, multi-party interviews in our study involved not just the agent and a single user, but a triad of bot, patient, and often informant. This setting created unique challenges for turn-taking and role clarity, echoing broader findings that voice assistants in shared spaces often face difficulties with overlapping talk and ambiguous addressees \cite{porcheron2018voice, meng2023multiuser}. To reduce confusion, the system used explicit visual cues to signal when the agent was speaking or listening, and it was designed to recognize the presence of both patient and informant, addressing the patient first while inviting the informant when needed. 

Beyond these technical strategies, role management was further complicated by the fact that, in our current setup, patients and informants were interviewed together in the same room. Prior work in caregiving contexts has similarly noted that co-present arrangements can suppress or shape what caregivers are willing to share \cite{10.1145/3491102.3517683, 10.1145/3491102.3502137, 10.1145/3613904.3642190}. In our study, informants sometimes withheld details or appeared cautious in front of the patient—for example, agreeing with the patient’s account even when they seemed to disagree, or holding back corrective information to avoid conflict. These same informants would often reveal conflicting information in the clinician interview, and indeed, some informants confided with study team members that they did not want to upset their loved one during the agent interview which they understood would not be used clinically, while they felt this must be done in the clinician interview to ensure the clinician received the information they needed for diagnosis and management. This was reflected in the informant responses accounting for a smaller share of turns in the agent interview (patient-to-informant turn ratio of nearly 3:1) compared to the clinician interview (1.5:1) To address this limitation, our planned phone call-based system (see Section \ref{subsec:system_overview}) separates the two interviews, allowing informants to contribute more candidly without the constraints of co-presence.

\subsection{Design Tradeoffs in Conversation Structure and Adaptivity}
Our results reinforce that conversation design is not a neutral technical choice, but a key determinant of both user experience and the quality of information elicited. Prompting strategy influenced the agent’s utterance-level behavior, producing trade-offs between conversational depth, systematic coverage, and user satisfaction. Sequential prompting fostered user engagement and improved diagnostic sensitivity through richer dialogue with follow-up and clarifying questions that extended beyond the scripted survey, but it also suffered from transition instability that reduced systematic coverage. Full-script prompting ensured comprehensive coverage of questions provided, but at the expense of conversational depth and user satisfaction. 

\textit{Sequential prompting} showed advantages over \textit{full-script prompting}. It made the interviews feel more natural with a higher user satisfaction and resulted in more clarifying questions (e.g., clarifying vague responses or requesting concrete examples when participants provided abstract descriptions). When there were misunderstandings, it was more likely to repair that misunderstanding, resulting in a more natural interview. The clarifying questions likely led to sequential prompting achieving higher symptom elicitation sensitivity and lower ambiguity rates, clearly demonstrating that design choices impact functional outcomes.  However, it also came with challenges—when the \textit{moved on to next topic} signal failed, the agent sometimes skipped content, lowering topic coverage, or became stuck on a topic, and in one case this caused the interview to become so long that the participant gave up. Therefore, it seemed to sacrifice breadth for greater depth. These issues are technically solvable, for example, by setting an upper bound on turns per topic to prevent the agent from stalling, or by restricting consecutive topic shifts to avoid unintentional skipping. The broader lesson, however, is that conversational agents for health must balance richness and completeness of elicitation with practical limits on time and participant effort. While \textit{full-script prompting} achieved greater coverage of topics, the agent frequently compressed multiple questions into single utterances, even when they belonged to different symptom domains. Therefore, the participant at times did not have the chance to answer the first question before the agent moved onto a second. This, as well as less frequent repair when misunderstandings occurred, reduced the natural flow of the conversation and likely led to the lower user satisfaction rates seen.

These findings surface broader implications for designing conversational agents in healthcare and other domains requiring comprehensive information gathering, clarifying ambiguous answers, and positive user experience. The challenge is achieving systematic coverage ensuring important topics are not overlooked while maintaining conversational naturalness that encourages detailed, thoughtful responses.

\subsection{Feasibility and Performance of Agent-Led History Taking}
Our pilot study demonstrates the feasibility of using conversational agents to elicit complex ADRD symptom narratives from older adults with cognitive impairment. Using a consensus-based gold standard derived from both agent and clinician transcripts, agent-led interviews achieved 83\% sensitivity and 100\% specificity for eliciting symptoms clearly present or absent. These performance metrics represent promising agreement with specialist interviews, but these results should be interpreted as proof-of-concept rather than evidence of equivalence to clinician assessment. It is possible the sensitivity is relatively low compared to specificity due to informants withholding information in the agent interview that they later revealed in the clinician interview as discussed above. Additionally, some symptoms may have been recognized only after reflection prompted by the initial interview.  The high specificity is a result of zero false positives in the agent interview, meaning participants never endorsed a symptom with the agent and subsequently denied it with the clinician. 

Agent interviews had lower not-discussed rates than clinician interviews across the 21 core symptoms. While these 21 symptoms were selected by two board-certified cognitive neurologists as symptoms that should ideally be assessed in every new dementia evaluation, there is still practice variation among dementia specialists and some may disagree with this list. Furthermore, the reality of clinical practice is that comprehensive one-hour new patient visits only afford 20-30 minutes of history taking; therefore, clinicians must be prudent about which questions to ask, focusing their time on those that are most likely to lead to the primary diagnosis and next steps in management. Regardless, it supports the agent's ability to perform time-intensive systematic questioning (median 46 minutes) compared to time-constrained clinical interviews (median 25 minutes for history-taking). From a design perspective, this suggests that conversational agents can handle comprehensive structured elicitation impractical in brief clinical encounters. While participants in our study demonstrated the willingness to engage in these lengthy interviews while maintaining overall high satisfaction, there may be a selection bias with these patients who were willing to participate in research and future design work will still need to balance comprehensiveness with participant time constraints. 

Importantly, ambiguity rates were comparable between agent and clinician interviews suggesting that when symptoms were elicited, agents obtained information of similar clarity to specialists. This is noteworthy given that older adults with cognitive impairment can find it challenging to describe complex symptoms, and supports our design choices to keep the participant engaged. Participants' survey responses highlighted the agent’s \textit{patience} and \textit{thoroughness} as positive qualities, echoing prior findings that conversational AI can impose less psychological burden than traditional assessments \cite{igarashi2024AIhumanstress}. Patients and informants provided longer responses to the agent than to the clinician as measured by word count per utterance, potentially due to patient and informant awareness of the time limitations of their clinic appointments. Such perceptions indicate that conversational agents can not only gather clinically relevant information but also foster trust, a prerequisite for adoption in sensitive healthcare contexts \cite{10.1145/3659625, 10.1145/3706598.3713839}. 

These findings are a first step in establishing feasibility for agent-based interviews with older adults with cognitive impairment and validate that thoughtful interaction design can support both comprehensive information gathering and positive user experience. While promising, this preliminary work leaves important questions about real-world deployment, optimal workflow integration, and comparative effectiveness to be addressed through future research..

\subsection{Limitations}
There are several limitations to consider. The fixed-order design with the agent always conducting the first interview may have further hindered agent performance relative to the clinician as patients and/or informants may have thought more about their symptoms through this interview or gathered additional information (e.g., asking family members about family history) that allowed more accurate or robust information during the second interview with the clinician. However, this was chosen to align with real-world workflows where the agent would always need to conduct the interview before the clinician. While we were able to demonstrate promising sensitivity and specificity using all symptoms combined, our sample size was too small to obtain a reliable sensitivity and specificity of each symptom. Another limitation is the variability in timing between the agent and clinician interview with some interviews up to 3 months apart. For slowly progressive disorders (i.e., ADRD), there are unlikely to be clinically meaningful changes during a 3 month period of time unless there is a rapidly progressive dementia; however, we acknowledge that some symptom evolution between interviews remains possible, and this limits how accurate the agent elicited symptoms can be. These factors limit causal claims about agent versus clinician symptom elicitation, though they do not undermine our core finding that appropriately designed conversational agents can feasibly engage older adults with cognitive impairment in extended diagnostic interviews.

A limitation arose from interviewing patients and informants together in the same room, a choice made to mirror typical clinician interviews. This likely reduced sensitivity as stated above. This disclosure asymmetry highlights a broader interaction design challenge: creating safe disclosure spaces when multiple stakeholders have competing social and informational needs. Critically, this work establishes feasibility and validates design choices for agent-based symptom elicitation—it does not validate clinical utility or deployment readiness. 

\subsection{Future Directions: Toward Clinically Useful Conversational Agents}
Current cognitive impairment screening tools (e.g., Montreal Cognitive Assessment (MoCA) and Mini Mental Status Exam (MMSE)), while valuable for initial detection, still create a significant downstream burden on an already overwhelmed healthcare system. Every positive screening requires a comprehensive clinical evaluation to confirm the result, rule out false positives, and then conduct time-intensive diagnostic evaluations identify the specific cause of cognitive symptoms (the underlying diagnosis) and initiate treatment. Pre-visit questionnaires in paper/PDF or online are commonly used to support this process, but suffer from low completion rates and practical considerations limit them to collection of yes/no responses with very few additional details \cite{shucard2022clinical}. While phone-based questionnaires have much higher completion rates \cite{nota2014differences}, their administration is limited by staff availability. Our agent-led interview system is designed to address this gap by functioning as a diagnostic support tool rather than a screener. Several open design challenges and research questions must be addressed before clinical deployment.

Clinicians interview informants and patients together due to clinic workflow limitations, but this agent interview system can be designed so that informants can be interviewed separately, leading to interaction design questions. How can conversational systems create safe disclosure spaces when multiple stakeholders have competing needs? Our planned phone-based deployment with separate patient and informant interviews is a first step towards addressing this, enabling access to people who cannot easily attend specialty clinics. However, this telephone system creates new interaction constraints, including no visual cues, potential audio quality issues, and handling interruptions in uncontrolled environments. Design research is needed to understand how conversational strategies should adapt to these conditions and how discrepancies between patient and informant reports can be presented to clinicians without privileging one voice over another. 

While the overall sensitivity and specificity of symptom elicitation was encouraging, the per symptom sensitivity and specificity were quite variable with large confidence intervals due to a low sample size. Clinical utility will require establishing which symptoms can be reliably gathered by conversational agents, ideally with very high sensitivity allowing clinicians to trust "absent" symptoms in order to focus time on "present" symptoms and/or clarifying ambiguous ones. Future analyses with a larger sample size will investigate whether additional symptoms elicited by the agent compared to the clinician benefit patient care, as it is possible clinicians correctly focused on gathering only the symptoms they needed for optimal patient care (i.e., appropriately less symptom coverage). If future research establishes clinical utility, integrating these tools into clinical workflows will require deliberate design choices and monitoring for safety. Rather than delivering raw transcripts, the system’s output should be concise, clinician-ready summaries (a "History of Present Illness") that highlight the symptoms most relevant to the patient's interaction. These summaries will need to entail a presentation of symptoms with examples and quotes to capture nuance, a timeline of onset and change, clear markers of not-discussed or ambiguous symptoms, and a note on informant–patient disagreements. Such structured summaries can support clinical care by flagging cases that require expedited evaluation, help clinicians target their questions more effectively, and shorten visits without sacrificing thoroughness. 

While no emergent safety issues were identified in our pilot study, this cannot be generalized to the real world setting as our participants understood this agent interview was solely for research purposes. Real-time monitoring systems would be needed for the conversational agents to automatically flag deviations from therapeutic objectives or identify emergencies. Design challenges include identifying concerning content (suicidal ideation, acute medical events) reliably enough to warrant clinician notification without excessive false alarms; creating escalation pathways that smoothly transition from automated conversation to human intervention; and maintaining participant trust when monitoring systems are transparent versus when they operate in the background. 

Finally, design questions emerge around data governance and patient agency. These interviews often contain sensitive health information. It will not be practical to conduct this research study's lengthy patient/informant consents in the clinic setting. What aspects of the consent process are most important to patients and their informants and therefore should be prioritized when deployed in the clinic? What controls should patients have over their interview data? How might systems transparently communicate data retention, access policies, and potential secondary uses? Addressing these open design challenges will require sustained collaboration between researchers, clinicians, and the older adults and informants who would use such systems.

%% file: section/conclusion.tex
Our study demonstrates the feasibility of using conversational agents to conduct semi-structured, voice-based diagnostic nterviews with older adults with cognitive impairment and their informants. Through iterative design refinement and empirical evaluation, we identified key interaction design considerations for voice interfaces in sensitive healthcare contexts. Agent-led interviews showed promising agreement with specialist assessment demonstrating that thoughtful interaction design can support comprehensive and accurate symptom elicitation. Participants valued the agent's patience, lack of time pressure, and question thoroughness. However, our findings also surface important design challenges: sequential prompting produced more natural conversations with higher satisfaction and more thorough questioning but unreliable symptom coverage relative to full-script prompting, informants withheld sensitive information when interviewed with patients, and comprehensive coverage sometimes created excessive interview length. Looking ahead, expanding to remote, phone-based settings and scaling participant enrollment will be critical for assessing clinical utility under real-world conditions. Taken together, these findings underscore both the promise and the design considerations required for developing clinically useful conversational agents in dementia care.

%% file: section/appendix.tex
\section{Appendix}
\subsection{Bot Interaction Behavior Annotation Guidelines}
\label{appendix:annotation_guideline}

\vspace{0.3cm}
\newcommand{\covered}{\textcolor{green}{\checkmark}} 
\newcommand{\notcovered}{\textcolor{red}{\textbf{X}}} 
\newcommand{\partialcovered}{\textcolor{orange}{\textbf{!}}} 
\newcommand{\surveybased}{\textcolor{blue}{\ding{109}}} 
\newcommand{\outofsurvey}{\textcolor{red}{\ding{55}}} 

\begin{tcolorbox}[width=\linewidth, colback=white, colframe=black, title=\textbf{Annotation Guideline for Clinical Interview Chatbot Conversation Quality Evaluation}]
\small
\setstretch{1.02}
\textbf{General Instructions} \\
\text{$\:\;$} - Annotate each chatbot utterance individually. \\
\text{$\:\;$} - Context is crucial: consider the patient’s and informant’s preceding responses. \\
\text{$\:\;$} - Use the provided clinical survey as the reference document. \\

\textbf{1. Survey Coverage (for each Topic Script)} \\
\textbf{Goal:} Determine whether the chatbot is systematically covering the required clinical survey items. \\
\textbf{Instructions:} \\
\text{$\:\;$} - Check whether the chatbot addressed the clinical survey item. \\
\text{$\:\;$} - If a survey item is conditional, verify whether the chatbot appropriately included or skipped it based on prior patient/informant responses. \\
\textbf{Mark as:} \\
\text{$\:\;$} - \textit{\covered $\;$ Covered:} The chatbot asked or addressed all items in the topic appropriately and accurately. \\
\text{$\:\;$} - \textit{\notcovered $\;$ Not Covered:} The chatbot did not ask any questions from the topic. \\
\text{$\:\;$} - \textit{\partialcovered $\;\;\:$  Partially Covered:} The chatbot addressed the questions in the topic only partially. \\

\textbf{2. Out-of-Survey Responses (for each Bot utterance)} \\
\textbf{Goal:} Identify and assess chatbot responses that do not correspond to the clinical survey. \\

\textbf{2a. Identification} \\
\textbf{Mark each chatbot response as:} \\
\text{$\:\;$} - \textit{\surveybased $\;$ Survey-Based:} The response aligns with a clinical survey item. \\
\text{$\:\;$} - \textit{\outofsurvey $\;$ Out-of-Survey:} The response is not traceable to any item in the survey. \\

\textbf{2b. Appropriateness} \\
\textbf{For each \textit{\outofsurvey $\;$ Out-of-Survey} response, assess clinical relevance:} \\
\text{$\:\;$} - \textit{\covered $\;$ Appropriate:} Adds rapport, clarifies information, addresses confusion, or supports the interview in a clinically acceptable way. \\
\text{$\:\;$} - \textit{\notcovered  $\;$ Inappropriate:} Irrelevant, distracting, confusing, or could bias the patient or harm rapport. \\
\textbf{Examples:} \\
\text{$\:\;$} - Chatbot says: “Okay, so you’ve been experiencing these problems for the last 3 years, where you have needed reminding of important events and also had to rely more on notes.” \\
\text{$\:\;\:\;$} - \textit{\outofsurvey $\;$Out-of-Survey} \\
\text{$\:\;\:\;$} - \textit{\covered $\;$ Appropriate} (summarizing to check understanding, without interrupting the interview) \\

\textbf{3. Antisocial / Impolite Responses (for each Bot utterance)} \\
\textbf{Goal:} Identify instances where the chatbot's responses are unempathetic, disrespectful, or offensive. \\
\textbf{Mark each chatbot response as:} \\
\text{$\:\;$} - \textit{\covered $\;$ Polite/Neutral:} Respectful and appropriate for elderly patients. \\
\text{$\:\;$} - \textit{\partialcovered $\;\:\:$ Impolite/Antisocial:} Includes any of the following: \\
\text{$\:\;\:\;\:\;\:\;$} - Lack of empathy or warmth \\
\text{$\:\;\:\;\:\;\:\;$} - Dismissiveness \\
\text{$\:\;\:\;\:\;\:\;$} - Insulting or judgmental tone \\
\text{$\:\;\:\;\:\;\:\;$} - Inappropriate humor or sarcasm \\
\text{$\:\;\:\;\:\;\:\;$} - Ageist or biased assumptions \\
\text{$\:\;\:\;\:\;\:\;$} - Responses that clearly lack tact or have insulting implications should also be marked “Impolite/Antisocial”  \\
\textbf{Examples:} \\
\text{$\:\;$} - “I see you’re having trouble speaking right now. Please speak clearly so that we can continue the interview.” → \textit{\partialcovered $\;$ Impolite} \\
\text{$\:\;$} - “It looks like there might be a microphone problem or another issue preventing me from understanding your answer. Would you please repeat your answer?” → \textit{\covered $\;$ Polite} \\

\textbf{4. Misunderstanding of Patient/Informant input (for each Bot utterance)} \\
\textbf{Goal:} Identify if the chatbot misunderstood the patient or misattributed speech between patient and informant. \\
\textbf{Mark each chatbot response as:} \\
\text{$\:\;$} - \textit{\covered $\;$ Understanding:} Correct Understanding \\
\text{$\:\;$} - \textit{\notcovered $\;$ Misunderstanding:} includes: \\
\text{$\:\;\:\;\:\;\:\;$} - Incorrect paraphrasing of patient/informant input \\
\text{$\:\;\:\;\:\;\:\;$} - Response to the wrong speaker (e.g., addresses the informant’s comment as if it were from the patient) \\
\text{$\:\;\:\;\:\;\:\;$} - Misinterpreting patient’s intent or meaning \\
\textbf{Examples:} \\
\text{$\:\;$} - Patient: “I don’t drive anymore.”
 Chatbot: “How often do you drive these days?” → \textit{\notcovered $\;$ Misunderstanding} \\
\text{$\:\;$} - Informant: “She forgets where she leaves things.”
 Chatbot: “You said you forget things often. Can you give an example?” → \textit{\notcovered $\;$ Misunderstanding} (misattributed speech) \\

\textbf{5. Failure to Repair Misunderstanding} \\
\textbf{Goal:} Evaluate whether the chatbot effectively corrects or redirects when a misunderstanding has occurred. \\
\textbf{Instructions:} \\
\text{$\:\;$} - For each previous misunderstanding by the chatbot OR misunderstanding by the patient, examine the chatbot’s next turn. \\
\textbf{Mark as:} \\
\text{$\:\;$} - \textit{\covered $\;$ Repaired:} The chatbot recognized and corrected the misunderstanding or appropriately redirected the conversation. \\
\text{$\:\;$} - \textit{\notcovered $\;$ Not Repaired:} The chatbot did not address or correct the issue, and the misunderstanding persists or worsens. \\
\textbf{Examples:} \\
\text{$\:\;$} - Patient: “Well my granddaughter was just telling me about what she was learning, she just started college, and I don’t remember any of that stuff even though I used to.”
 Chatbot: “I see. What about forgetting things that happened more recently? Have you had trouble keeping track of your schedule or remembering things that happened in the past few years?” → \textit{\covered $\;$ Repaired} \\
\text{$\:\;$} - Patient: “Well my granddaughter was just telling me about what she was learning, she just started college, and I don’t remember any of that stuff even though I used to.”                                      Chatbot: “Do you find that you have to rely more on notes than you used to in your daily life?” → \textit{\notcovered $\;$ Not Repaired} (failed to clarify) \\

\textbf{6. No Opportunity to Answer} \\
\textbf{Goal:} Identify instances where the chatbot poses a question appropriately but does not provide space for the patient to respond before continuing. \\
\textbf{Instructions:} \\
\text{$\:\;$} - Asks a relevant and clearly phrased question, but immediately follows it with another question, statement, or topic shift. \\
\text{$\:\;$} - Does not give the patient a natural conversational opportunity to respond. \\
\textbf{Mark as:} \\
\text{$\:\;$} - \textit{\covered $\;$ Opportunity Provided:} The chatbot gives a clear opportunity (either by pausing or awaiting input) after asking a question. \\
\text{$\:\;$} - \textit{\notcovered $\;$ No Opportunity to Answer:} - The chatbot continues speaking too quickly or shifts topics without allowing for a pause or patient input. \\
\textbf{Examples:} \\
\text{$\:\;$} - Chatbot: “Can you tell me more about that?” [waits or includes pause] → \textit{\covered $\;$ Opportunity Provided} \\
\text{$\:\;$} - Chatbot: “How has your memory been lately? Have you been sleeping well?” → \textit{\notcovered $\;$ No Opportunity to Answer}
\end{tcolorbox}
\vspace{-0.5cm}
\captionof{figure}{Annotation guideline used for chatbot conversation quality evaluation.}
\label{fig:guideline-annotation}